# Giant magnetoresistance and structure of electrodeposited Co/Cu multilayers: the influence of layer thicknesses and Cu deposition potential

N. Rajasekaran[a,b], J. Mani[b], B.G. Tóth[a], G. Molnár[c], S. Mohan[b], L. Péter[a], I. Bakonyi[a,*]

[a]*Wigner Research Centre for Physics, Hungarian Academy of Sciences. H-1121 Budapest, Konkoly-Thege út 29-33, Hungary*

[b]*Central Electrochemical Research Institute. Karaikudi-630006, Tamil Nadu, India*

[c]*Institute for Technical Physics and Materials Science,Hungarian Academy of Sciences. H-1121 Budapest, Konkoly-Thege út 29-33, Hungary*

**Abstract** – The giant magnetoresistance (GMR) and structure was investigated for electrodeposited Co/Cu multilayers prepared by a conventional galvanostatic/potentiostatic pulse combination from a pure sulfate electrolyte with various layer thicknesses, total multilayer thickness and Cu deposition potential. X-ray diffraction (XRD) measurements revealed superlattice satellite reflections for many of the multilayers having sufficiently large thickness (at least 2 nm) of both constituent layers. The bilayer repeats derived from the positions of the visible superlattice reflections were typically 10 – 20% higher than the nominal values.The observed GMR was found to be dominated by the multilayer-like ferromagnetic (FM) contribution even for multilayers without visible superlattice satellites. There was always also a modest superparamagnetic (SPM) contribution to the GMR and this term was the largest for multilayers with very thin (0.5 nm) magnetic layers containg apparently a small amount of magnetically decoupled SPM regions. No oscillatory GMR behavior with spacer thickness was observed at any magnetic layer thickness. The saturation of the coercivity as measured by the peak position of the MR(H) curves indicated a complete decoupling of magnetic layers for large spacer thicknesses. The GMR increased with total multilayer thickness which could be ascribed to an increasing SPM contribution to the GMR due to an increasing surface roughness, also indicated by the increasing coercivity. For multilayers with Cu layers deposited at more and more positive potentials, the $GMR_{FM}$ term increased and the $GMR_{SPM}$ term decreased. At the same time, a corresponding reduction of surface roughness measured with atomic force microscopy indicated an improvement of the multilayer structural quality which was, however, not accompanied by an increase of the superlattice reflection intensities. The present results underline that whereas the structural quality as characterized by the surface roughness generally correlates fairly well with the magnitude of the GMR, the microstructural features determining the amplitude of superlattice reflections apparently do not have a direct influence on the GMR.

*Corresponding author. E-mail: bakonyi.imre@wigner.mta.hu

## Introduction

Due to the large giant magnetoresistance (GMR) effect observed in physically deposited Co/Cu multilayers [1-3], a lot of efforts have been devoted to the study of GMR also on electrodeposited (ED) Co/Cu multilayers (for detailed references, see a recent review [4]). A variety of baths have been used for the preparation of ED Co/Cu multilayers [4], the simplest one containing merely $CoSO_4$ and $CuSO_4$. Over the last two decades, numerous reports have been published on studying the GMR characteristics of ED Co/Cu multilayers from the pure sulfate bath (containing at most some buffering agents) [5-15]. In this list of references, we have included only those works from the much larger number of reports [4] in which the ED Co/Cu multilayers were prepared from the sulfate bath at or close to the electrochemically optimized Cu deposition potential $E_{Cu}^{EC}$ [5,10,16] where neither Co dissolution, nor Co codeposition can occur during the Cu pulse. These features ensure that the actual layer thicknesses will be fairly close to the nominal values and that the spacer layer will not contain magnetic Co atoms.

By looking at former reports [5-15], it can be established that in most cases not completely systematic studies on layer thicknesses have been carried out for ED Co/Cu multilayers from the sulfate bath. For example, there were several studies of the GMR dependence on layer thicknesses, where one of the layer thicknesses was fixed and the other layer thickness was only varied. Even in these cases, the total multilayer thickness covered a very wide range, in some cases up to the micrometer scale. It has been known [17], however, that the roughness increases strongly with increasing deposit thickness and recent reports [18-20] have shown, on the other hand, that the GMR correlates sensitively with the roughness of ED multilayers. Although in one of our early previous reports [8], we made a detailed study of GMR as a function of both kinds of layer thickness but in this particular case fairly thick (1.7 μm) ED Co/Cu multilayers were prepared on a mechanically polished and, thus, very rough Ti foil substrate so the above mentioned roughness problem may have been pertinent also here.

Another deficiency of these previous studies in several cases could have been that electrodeposition was performed in an open cell geometry providing space for edge effects and the cathode position was often vertical. When measuring GMR on multilayer films with the usual van der Pauw geometry [21], these effects do not become easily evident but applying a four-point-in-line method for GMR measurements on narrow multilayer strips can clearly reveal them [22,23]. For eliminating these deleterious effects and preparing laterally homogeneous deposits, a tubular cell has been designed [24,25] in which the cathode is at the bottom of the cell with an upward looking deposition area filling the whole cross section of the cell.

It is evident from the foregoing discussion that in previous studies the preparation of the ED Co/Cu multilayers from the sulfate bath was not optimal in every respect (and the same holds true also for studies using different bath formulations [4]). It appeared, therefore, worthwhile to carry out a systematic study of GMR on ED Co/Cu multilayers from the pure sulfate bath under well-controlled conditions which include (i) the use of a smooth Si/Cr/Cu substrate obtained by evaporating nanometer-scale Cr and Cu layers on a Si wafer, (ii) the deposition of a multilayer with a constant total thickness of only 300 nm, (iii) the use of an electrochemical cell ensuring very good lateral homogeneity [24,25] and (iv) the application of an optimized galvanostatic/potentiostatic (G/P) pulse combination [8,16,25].

Recently, we have already presented some results on ED Co/Cu multilayers prepared from a pure sulfate bath [26]. In that work, first a series of multilayers with constant magnetic layer thickness and varying Cu layer thickness was prepared by a G/P pulse combination at the electrochemically optimized Cu deposition potential $E_{Cu}^{EC}$. In a second series, ED Co/Cu multilayers were prepared in a manner that in each cycle a bilayer of Co(2.0 nm)/Cu(6.0 nm) was first prepared from which a small fraction of the Cu layer was dissolved by a third galvanostatic anodic pulse of various lengths in order to achieve the same nominal spacer thicknesses as in the first series. This study was intended to introduce a method for controlling the microstructure of the Cu layer which definitely has a strong influence on the GMR.

In the present work, we will describe further results on the GMR of ED Co/Cu multilayers prepared from a pure sulfate bath under the above described refined conditions by varying both the magnetic and non-magnetic layer thicknesses. Since a recent study on ED Ni-Co/Cu multilayers [20] has indicated that the electrochemically optimum potential does not necessarily correspond to the GMR optimum, we wanted now to explore the influence of Cu deposition potential on the GMR also for ED Co/Cu multilayers.

Furthermore, X-ray diffraction (XRD) and atomic force microscopy (AFM) measurements have also been performed in order to characterize the structure and the surface roughness of these multilayer samples.

## Experimental

*ED Co/Cu multilayer preparation and characterization.* — The basic electrodeposition conditions were those applied in our recent work [26] where magnetic/non-magnetic Co/Cu multilayers were prepared from an aqueous electrolyte containing 1 M $CoSO_4$ and 0.025 M $CuSO_4$. The multilayer electrodeposition was performed on a Si(100)/Cr(5 nm)/Cu(20 nm) substrate where the Cr adhesive and Cu seed layers were obtained by evaporation. Electrodeposition was carried out in a tubular cell [24,25] at room temperature in which the substrate was at the bottom of the cell with upward looking cathode surface area of about 7.5 mm by 20 mm. This arrangement ensures a lateral homogeneity of the deposits and helps to avoid edge effects.

For the present study, most Co/Cu multilayers were electrodeposited by the conventional two-pulse plating in the mixed galvanostatic/potentiostatic (G/P) deposition mode [25] in which the magnetic layer is deposited by controlling the deposition current (G mode), whereas the non-magnetic layer (pure Cu) is deposited by controlling the deposition potential (P mode). The magnetic layer deposition was carried out at a fixed cathodic current density amplitude of -50 mA/cm$^2$. According to a detailed analysis on a large set of multilayers prepared under similar conditions [27] for such multilayers, the current efficiency during the magnetic layer deposition is almost unity. Therefore, the nominal magnetic layer thicknesses were calculated on this basis from Faraday’s law by taking into account the length of the G pulse. For the Cu layer deposition, the deposition potential was electrochemically optimized [8,16,26] to be $E_{Cu}^{EC} = -600$ mV with respect to a saturated calomel electrode (SCE). The steady-state diffusion-limited Cu deposition current density was -1.4 mA/cm$^2$ [26]. The nominal Cu layer thickness was set by measuring the charge passed through the cell and by using Faraday’s law under the usual assumption of 100 % current efficiency for Cu deposition at the limiting current density. From the ratio of the diffusion-limited Cu deposition current density (P pulse) to the current density used for Co-layer deposition (G pulse), it can be estimated (see p. 135 and Fig. 16 in Ref. 4) that the Cu content of the magnetic layer is about 2.8 at.% when producing the Co/Cu multilayers with the G/P pulse sequence.

According to the aims described in the Introduction, four sets of ED Co/Cu multilayers as summarized in Tables 1 to 4 were prepared for the present study on the basis of the above parameters. First, three ED Co/Cu multilayer sets with various Co and Cu layer thicknesses (while keeping the total multilayer thickness at 300 nm, see Tables 1 and 2) as well as with various total multilayer thicknesses from 50 nm to 300 nm (while keeping the individual layer thicknesses constant at Co(1.1 nm)/Cu(4.0 nm) and Co(2.0 nm)/Cu(4.0 nm), see Table 3) were prepared. The fourth set of samples consisted of Co/Cu multilayers with two combinations of fixed layer thicknesses (1.1 nm or 2.0 nm for Co and 4.0 nm for Cu) and a fixed total multilayer thickness of 300 nm whereby the Cu deposition potential was varied from $E_{Cu}^{EC}$ = -600 mV to -250 mV, to a value where a significant dissolution of the magnetic layer [25] is expected to occur (see Table 4).

It should be noted that for G/P multilayers deposited at $E_{Cu}^{EC}$, the actual layer thicknesses correspond fairly well to the nominal values as determined above which was evidenced from detailed XRD studies [13,14]. However, for multilayers for which the applied Cu deposition potential is more positive than $E_{Cu}^{EC}$, the actual magnetic layer thickness will be smaller and the Cu layer thickness larger than the corresponding nominal values. The reason of this layer thickness deviation is the partial dissolution of Co atoms from the last magnetic layer during the Cu layer deposition pulse. Since the Co dissolution is associated with a positive (anodic) current contribution, an equivalent amount of Cu atoms has to be deposited, regarding that a charge count is used to control the Cu layer thickness. By previous detailed compositional studies of such multilayers [25,28], it could be established that layer thickness changes as high as 1.4 nm can occur at Cu deposition potentials much less negative than $E_{Cu}^{EC}$.

It has to be noted that the capacitive nature of the electrode surface also plays a role in the occurrence of a difference between nominal and real layer thicknesses. Even at optimized conditions, there remains a capacitive transient at the beginning of each current pulse. This is well evidenced by various research groups for the deposition pulse of the more noble element (see, e.g., Figs 7, 9 and 10 of Ref. 16 and Figs. 3, 4, and 5 of Ref. 29). Due to a capacitive current contribution during the deposition pulse of the magnetic layer, the thickness of the magnetic layer becomes smaller than the value corresponding to the total charge counted (this is because the capacitive contribution has the same sign as the Faradaic current). On the other hand, the capacitive contribution during the Cu layer deposition pulse results in a larger Cu layer thickness with respect to that derived from the charge count (in this case, the capacitive charging contribution has the opposite sign as the Faradaic current). However, this capacitive effect is of rather secondary importance only beside the proper choice of the copper deposition potential.

For G/P multilayers with the Cu layer deposited at potentials less negative than $E_{Cu}^{EC}$, the bilayer thickness remains unchanged with respect to the multilayers prepared with Cu layers at $E_{Cu}^{EC}$ . In general, the bilayer lengths derived from either TEM or XRD were in fairly good agreement with the nominal values, the experimental data being typically 10 to 20 % higher [13,14,26-28].

In order to trace out the layer thickness changes in the fourth set of multilayers, the overall composition of the multilayers also had to be determined. This analysis was carried out by energy dispersive X-ray fluorescence (EDXRF) method with an XGT-HORIBA-5200 instrument.

Structural information was obtained by using XRD and Lorentzian curves were fitted to the background-corrected XRD diffraction patterns to determine the peak positions. Where multilayer satellite reflections were observed, the bilayer thickness was calculated from the satellite peak positions.

The root-mean-square surface roughness ($R_q$) of the deposited multilayers was determined by using atomic force microscopy with an Agilent Technologies 5500 instrument.

*Magnetoresistance.* — The magnetoresistance measurements were performed at room temperature with the four-point-in-line method in magnetic fields $H$ between -8 kOe and +8 kOe in the field-in-plane/current-in-plane geometry. Both the longitudinal (LMR) and the transverse (TMR) magnetoresistance (field parallel to current and field perpendicular to current, respectively) components were recorded for each sample. The following formula was used for calculating the magnetoresistance ratio: $\Delta R/R_0 = [R(H) - R_0]/R_0$ where $R(H)$ is the resistance in the magnetic field $H$ and $R_0$ is the resistance maximum value around $H = 0$. A shunting-effect correction due to the metallic underlayers on the substrate was done on the measured MR data by using the measured values of the zero-field resistivity of both the substrate and the substrate/multilayer stack [18]. It turned out that this correction is negligible for multilayers with 300 nm total thickness (i.e., for most of the samples studied) so an actual correction had to be performed only for multilayers with total thickness less than 300 nm. The measured field dependence of the magnetoresistance, MR(H) was decomposed according to a standard procedure [23] into ferromagnetic (FM) and superparamagnetic (SPM) contributions of the GMR.

For most multilayers investigated, the measured field dependence of the LMR and TMR components was very similar and both exhibited negative values indicating a clear GMR effect. Due to the unavoidable presence of an anisotropic magnetoresistance (AMR) contribution [30,31] in the magnetic layers for which LMR > 0 and TMR < 0 (and, by definition, AMR = LMR - TMR), the measured TMR values were slightly higher (by about 0.5 to 1 %) than the LMR data obtained for the same multilayer. From the measured LMR and TMR data, the saturation value of the isotropic GMR contribution can be obtained as $GMR_{is} = |(1/3)\,LMR + (2/3)\,TMR|$ [4]. For the sake of clarity of the data presented, the isotropically averaged $GMR_{is}$ data will only be displayed in most cases.

## Results on multilayers with Cu layers deposited at $E_{Cu}^{EC}$ = -600 mV

*Structural study by XRD.* — The XRD patterns around the main fcc-(111) multilayer reflection are shown for a series with various Co layer thicknesses in Fig. 1a and for a series with various Cu layer thicknesses in Fig. 1b.

For a multilayered structure with sufficient coherence along the growth direction, the appearance of superlattice satellite reflections can be expected [32] as can clearly be seen for $t_{Co}$ = 2.0 nm and 3.0 nm in Fig. 1a. If one of the layers is too thin, the superlattice reflection may be disrupted due to the uneven layer growth [13]. This is definitely the case for $t_{Co}$ = 0.5 nm whereas for $t_{Co}$ = 1.1 nm, a sign of superlattice reflections can still be recognized. The distance of satellites from the main reflection is inversely proportional to the bilayer length $\Lambda = t_{Co} + t_{Cu}$ [32,33]. The dashed lines indicate a corresponding evolution of the satellite peak positions.

For the multilayers shown in Fig. 1(b), both kinds of layer thicknesses are sufficiently thick and for this reason all the multilayers exhibit clear satellite reflections. The expected evolution of the satellite peak position is again indicated by the dashed lines. For this series, the main peak position shift can also be clearly observed with the help of the central vertical dashed line which corresponds to an increasing average lattice parameter for thicker Cu layers.

The XRD patterns for multilayers with constant layer thicknesses but varying total multilayer thickness are presented in Fig. 2. For the multilayer series [Co(1.1 nm)/Cu(4 nm)], the satellite reflections are visible for the 300 nm thick multilayer only whereas the satellites are completely missing for thinner multilayers. For the multilayer series [Co(2 nm)/Cu(4 nm)] with varying total multilayer thickness, the corresponding XRD patterns are shown in Fig. 2b. In agreement with the general trend discussed above in connection with Figs. 1 and 2, due to the sufficiently thick Co layer, most of the multilayers in this series exhibit satellite reflection (at least on the low-angle side of the main peak).

Where it was possible, the satellite peak position was determined by fitting the observed XRD patterns shown in Figs. 1 and 2 in order to deduce the experimental bilayer length $\Lambda_{XRD}$ and the results are summarized in Fig. 3 for all four series. It can be established that the $\Lambda_{XRD}/\Lambda_{nom}$ ratio where $\Lambda_{nom}$ is the bilayer length obtained from the nominal layer thicknesses is typically 10-20 % higher than 1. This corresponds to the general trend summarized previously on ED Co/Cu multilayers [26], including also data derived with the help of a more sophisticated full-profile fitting procedure [13-14] or even direct cross-sectional TEM imaging [28].

*Magnetoresistance.* — The MR(H) curves measured up to H = 8 kOe for the investigated multilayer series are characterized by the two types shown in Fig. 4. For cobalt layers as thin as 0.5 nm, for all Cu layer thicknesses usually a non-saturating MR(H) was obtained as shown in Fig. 4(a) for the multilayer $[Co(0.5\,nm)/Cu(5.0\,nm)]_{300nm}$. This can be ascribed to the presence of a significant fraction of SPM particles within the magnetic layers [23]. Such a non-saturating character of the MR(H) curves sometimes appeared also if both layer thicknesses were not very thick (e.g., $[Co(1.1\,nm)/Cu(2.0\,nm)]_{300nm}$). A Langevin-fitting of the high-field section of the MR(H) curves enables the separation of the $GMR_{FM}$ and $GMR_{SPM}$ contributions from the measured data [23] as indicated in Fig. 4. For sufficiently thick Co and Cu layers, the MR(H) curves exhibited significantly less non-saturating character as demonstrated in Fig. 4(b) for the $[Co(3.0\,nm)/Cu(5.0\,nm)]_{300nm}$ multilayer. In this case, the total observed magnetoresistance at high field is dominated by a $GMR_{FM}$ contribution as revealed in Fig. 4b.

The GMR data were measured for four $[Co(t_{Co})/Cu(t_{Cu})]_{300nm}$ multilayer series (Tables 1, 2 and 3) up to 8 kOe. By performing the Langevin-fitting for the measured MR(H) data, the saturation values of $GMR_{FM}$ and $GMR_{SPM}$ contributions as well as of $GMR_{is}$ were determined for all multilayers. The obtained $GMR_{is}$ data are displayed in Fig. 5a as a function of the spacer layer thickness $t_{Cu}$ and in each series with a constant magnetic layer thickness $t_{Co}$ as indicated in the legend. The evolution of the saturation values of the $GMR_{FM}$ and $GMR_{SPM}$ terms is shown in Fig. 5b.

For the series with $t_{Co}$ = 1.1, 2.0 and 3.0 nm, the evolution of GMR with spacer thickness corresponds well to previous observations [5,8,9,11,12] in that for low $t_{Cu}$ values the $GMR_{FM}$ component increases and then for larger spacer thicknesses, it saturates which is then followed by a decrease of the GMR. The latter feature can be ascribed to a simple dilution effect (the smaller number of magnetic/non-magnetic interfaces per unit thickness leads to fewer spin-dependent scattering events contributing to the GMR effect). The $GMR_{FM}$ data in Fig. 5b support previous conclusion [12] that there is no oscillatory GMR in ED Co/Cu multilayers due to the absence of an alternating antiferromagnetic coupling between adjacent magnetic layers.

The series with $t_{Co}$ = 0.5 nm does not fit into the scheme described above for the $GMR_{FM}$ term since the GMR in this series is smaller than in the series with $t_{Co}$ = 1.1 nm. Explaining this deviation requires a more detailed analysis based on a closer look at the field dependence of the magnetoresistance [23]. In the case of the three series with $t_{Co}$ = 1.1, 2.0 and 3.0 nm, the $GMR_{SPM}$ component is fairly small with respect to $GMR_{FM}$ and is typically around 1 % whereas the $GMR_{FM}$ term was as high as 8 % in some cases (Fig. 5b). This implies that whereas there is always a small fraction of the magnetic layers which exhibits SPM behavior, the majority of the magnetic layers has a predominantly ferromagnetically behaving character by forming continuous layers and the same holds true even if these layers are discontinuous but fully percolating in the layer plane. This is because a dominating $GMR_{FM}$ term can only arise from spin-dependent scattering events for electron pathways through a NM spacer between adjacent FM layers with non-aligned magnetizations. The $GMR_{SPM}$ term originates from electron scatterings when conduction electrons travel between a FM and a SPM region [23]. As discussed in previous reports [23,34], due to the layered structure, the actual SPM fraction of the magnetic layers is certainly smaller than the ratio of the $GMR_{SPM}$ and $GMR_{FM}$ terms. By contrast, the $GMR_{SPM}$ term is as high as 1/3 of the $GMR_{FM}$ term for the series with $t_{Co}$ = 0.5 nm (Fig. 5b). This is a clear indication that in this series the magnetic layer is highly discontinuous and a non-negligible fraction of it is in the form of magnetically decoupled regions exhibiting SPM character. The non-uniform lateral distribution of the magnetic material in the plane of the nominally very thin magnetic layer results in a reduction of the area of the FM/NM interfaces, which are the source of spin-dependent scattering events leading a multilayer-type GMR. These features result then in a reduced $GMR_{FM}$ term for the series with $t_{Co}$ = 0.5 nm and, thus, a smaller total $GMR_{is}$ value (Fig. 5a) with respect to the series with $t_{Co}$ = 1.1 nm.

Figure 5c shows the peak positions ($H_p$) of the MR(H) curves as a function of the Cu layer thickness in each series discussed above with a constant value of the magnetic layer thickness. The value of $H_p$ roughly corresponds to the coercive field $H_c$ of the magnetic layers. The data shown exhibit the same trend as reported previously [12] in that the $H_p$ values for large $t_{Cu}$ approach saturation. This was interpreted [12] by the fact that sufficiently thick spacer layers are already completely continuous and, thus, magnetically decouple the magnetic layers from each other which can, therefore, behave as individual thin layers. If this decoupled state is achieved for thick Cu layers, the coercive field is not expected to change any longer as we can indeed observe in Fig. 5c in agreement with Ref. 12.

After measuring the MR(H) data for the multilayer series [Co(1.1 nm)/Cu(4.0 nm)] and [Co(2.0 nm)/Cu(4.0 nm)] with varying total multilayer thickness (Table 3), Langevin fitting was carried out to obtain the saturation values of the $GMR_{FM}$ and $GMR_{SPM}$ contributions. These values were then corrected for the shunting effect of the substrate where it was necessary and are plotted on Figure 6 in order to see the evolution of GMR with total multilayer thickness.

Apart from 50 nm total multilayer thickness, the $GMR_{FM}$ values are larger for the series with $t_{Co}$ = 1.1 nm with respect to the series $t_{Co}$ = 2.0 nm (Fig. 6b) and this can be explained by the higher number of FM/NM interfaces in the former which leads to higher $GMR_{FM}$ in the first series. Since the magnetoresistance data have been corrected for the shunting effect of the substrate, this effect can be ruled out as the source of the observed increase of $GMR_{is}$ with total multilayer thickness (Fig. 6a). As Fig. 6b shows, the increase of the total GMR can be well understood in terms of the observed an increase of the $GMR_{SPM}$ component (Fig. 6b). With increasing total thickness, the multilayers usually exhibit a surface roughening [17,19], and this may well be a reason for the large increase of the $GMR_{SPM}$ term. Namely, it was suggested [35 that a possible mechanism of SPM region formation in multilayers is an increase in surface roughness.

The peak position values ($H_p$) of the MR(H) curves were found to increase with total multilayer thickness (Fig. 6c). The increasing roughness with total multilayer thickness is an evident explanation also for this latter observation.

### Results on multilayers with Cu layers deposited at $E_{Cu}$ > $E_{Cu}^{EC}$ = -600 mV

*Composition analysis and layer thickness changes.* — The overall composition of the two multilayer series $[Co(1.1\ nm)/Cu(4.0\ nm)]_{300nm}$ and $[Co(2.0\ nm)/Cu(4.0\ nm)]_{300nm}$, which were prepared with various Cu deposition potentials (Table 4), was measured while being on their Si/Cr(5 nm)/Cu(20 nm) substrates. The measured Cu contents are displayed in Fig. 7 as a function of the Cu deposition potential for the two series. The Cu content shows a fairly monotonous increase of about 7 % for both series (there is a corresponding decrease in Co content not shown here).

At the electrochemically optimized potential $E_{Cu}^{EC}$ = -600 mV, the actual layer thicknesses are expected to be equal to the nominal values. Under this assumption and by taking into account the 2.8 at.% Cu content in the magnetic layer, we obtain 79.0 at.% Cu and 67.6 at.% Cu for the multilayers Co(1.1 nm)/Cu(4.0 nm)]$_{300nm}$ and Co(2.0 nm)/Cu(4.0 nm)]$_{300nm}$, respectively, when prepared at -600 mV Cu deposition potential. If we also take into account the substrate Cu layer contribution, we arrive at 80.6 at.% Cu and 69.6 at.% Cu which should be compared to the measured values (84.3 and 77.5 at.% Cu, respectively) for these two multilayers. The measured data reflect well the differences due to the different Co layer thicknesses for the two samples although they are larger than the expected values by about 4 and 8 %, respectively. It has been our general experience that there is a relatively large uncertainty of the chemical composition analysis in the Co-Cu binary system by the SEM technique, usually beyond the typical error (of the order ±1 at.%) achievable for other element combinations. We believe this is the reason for the large discrepancy between the measured and expected values.

For the multilayers Co(1.1 nm)/Cu(4.0 nm)]$_{300nm}$ and Co(2.0 nm)/Cu(4.0 nm)]$_{300nm}$ deposited at a Cu deposition potential more positive than -600 mV, we can expect a partial dissolution of the magnetic layer and a corresponding deposition of an excess amount of Cu according to the discussion in the Experimental section. This can explain the observed monotonous increase of the Cu content as the Cu deposition potential changes from -600 mV to -250 mV. By taking the observed increase of the measured Cu content at the most positive deposition potential (-250 mV), we can estimate an increase of the Cu layer thickness by about 0.35 nm for 1.1 nm Co layer thickness and by 0.50 nm for 2.0 nm Co layer thickness (and, evidently, a corresponding decrease of the Co layer thicknesses). The layer thickness changes due to the various degree of Co dissolution as the Cu deposition potential is varied can be assumed to scale with the linear variation of the overall Cu content.

*AFM study of surface roughness.* — The surface roughness has been studied by AFM for the multilayer series [Co(2 nm)/Cu(4 nm)]$_{300nm}$ with Cu layers deposited at potentials from -600 mV to -250 mV and the measured roughness values are displayed in Fig. 8. There is a very drastic reduction of the roughness from -600 mV to -400 mV. Since at these potentials, the Co layer is being partially dissolved during the Cu pulse, this leads to a roughening as demonstrated by a previous study [18] on very thin ED Co/Cu multilayers. A particularly strong smoothening effect due to the magnetic layer dissolution at Cu deposition potentials more positive than $E_{Cu}^{EC}$ was reported also for ED Ni-Co/Cu multilayers [20].

The detailed AFM results on the present ED Co/Cu multilayers have revealed (Fig. 8) that at excessive Co dissolution (Cu deposition potentials more positive than -400 mV, the surface roughness increases again although it still remains below the roughness vale obtained at $E_{Cu}^{EC}$.

*Structural study by XRD.* — The XRD patterns around the main fcc(111) multilayer reflection are shown in Fig. 9 for the two multilayer series [Co(1.1 nm)/Cu(4.0 nm)]$_{300nm}$ and [Co(2.0 nm)/Cu(4.0 nm)]$_{300nm}$, both prepared at various Cu deposition potentials (Table 4). As discussed above, there is a small increase of the Cu layer thickness towards more positive Cu deposition potentials and, therefore, due to the larger lattice parameter of Cu, a slight shift of the multilayer fcc(111) main peak position to lower angles should occur as actually can be observed in Fig. 9.

For the multilayer series [Co(1.1 nm)/Cu(4.0 nm)]$_{300nm}$, very faint satellite reflections are only visible (and even here the larger lower-angle satellites only). This corresponds to the tendency discussed in connection with Fig. 1 that if one of the constituent layers is small, the satellite intensity is small or the satellites may even disappear (note that the Co layer thickness is even further reduced slightly for potentials more positive than $E_{Cu}^{EC}$). In the multilayer series [Co(2.0 nm)/Cu(4.0 nm)]$_{300nm}$, clear satellites on both sides of the fcc(111) main peak can be observed for $E_{Cu}$ = -600 mV. The satellite intensities then progressively diminish towards more positive Cu deposition potentials.

If we compare the surface roughness (Fig. 8) and satellite intensity (Fig. 9b) evolutions with Cu deposition potentials for the multilayer series [Co(2.0 nm)/Cu(4.0 nm)]$_{300nm}$, it can be established that whereas both roughness and satellite intensity are usually taken as an indicator of structural quality of multilayers, these two parameters are not correlated here with each other. A similar conclusion has already been achieved above in connection with the multilayer series having varying total multilayer thickness. For the Ni-Co/Cu system, on the other hand, the satellite intensity clearly increased with improving surface roughness [20]. Apparently, the interface structure to which the satellite intensity is sensitive changes differently depending on whether one has to do with one magnetic layer component (Co/Cu multilayers) or with two components in the magnetic layer (Ni-Co/Cu).

In contrast to the shift of the main multilayer XRD peak position with varying Cu deposition potential, the positions of the satellite peaks with respect to the main peak are expected to be unchanged since the bilayer thicknesses are not modified due to the Co dissolution process. The XRD patterns in Fig. 9 roughly correspond to this for both series.

From the observed satellite peak positions of Figs. 9a and 9b, the experimental bilayer length $\Lambda_{XRD}$ was determined where it was possible and the results are summarized in Fig. 9c. Similarly to the previous series above, the $\Lambda_{XRD}/\Lambda_{nom}$ ratio was again typically 10-20 % higher than 1.

*Magnetoresistance.* — The evolution of the measured $GMR_{is}$ at the maximum applied field (8 kOe) for the multilayer series $[Co(1.1\ nm)/Cu(4.0\ nm)]_{300nm}$ and $[Co(2.0\ nm)/Cu(4.0\ nm)]_{300nm}$ is shown in Fig. 10a as a function of the Cu deposition potential $E_{Cu}$. At a Cu deposition potential of -600 mV, in agreement with Fig. 5, the GMR is larger for multilayers with 1.1 nm Co layer thickness than for multilayers with 2.0 nm thick Co layers. However, when $E_{Cu}$ is more positive than the electrochemically optimized potential $E_{Cu}^{EC}$ = -600 mV, i.e., when a partial dissolution of the Co layer occurs during the Cu layer deposition, the GMR becomes smaller for the thinner Co layers. This feature must be a consequence of the dissolution process which attacks a larger relative fraction of the thinner Co layer than for the thicker one. This must also be the reason for the peculiar, non-monotonic evolution of GMR for $t_{Co}$ = 1.1 nm since it is hard to assess the influence of the dissolution process for the thin Co layer being itself at the borderline of continuity already.

For the thicker Co layer, the observed GMR shows a clear increase towards more positive potentials where the dissolution process becomes stronger and stronger. As discussed above, the layer thickness changes can be estimated to be at most 0.5 nm in both series. Since the repeat period remains unchanged for any Cu deposition potential, only the slight increase of the Cu layer thickness could give a contribution to the increase of GMR as $E_{Cu}$ varies from -600 mV to -250 mV. However, with reference to Fig. 5 where the Cu layer thickness dependence of GMR was displayed, we can assess that the increase of the Cu layer thickness cannot give a significant contribution to the observed large GMR increase which is as high as 4 % for 2.0 nm thick Co layers. Therefore, the major cause of the GMR increase with $E_{Cu}$ must be due to the surface smoothening as a consequence of the dissolution process as was already observed also for electrodeposited Ni-Co/Cu multilayers [20]. Along the same line, smaller degree of the GMR increase for $E_{Cu}$ values more positive than -400 mV (Fig. 10a) might be connected with the rise of the surface roughness in this potential range (Fig. 8).

The results of the decomposition of the total GMR into FM and SPM contributions are shown in Fig. 10b. For both series, the major contribution to the GMR is from the multilayer GMR mechanism ($GMR_{FM}$). The SPM contribution is fairly small and does not change too much for the Cu deposition potentials investigated. Therefore, the observed significant increase of the $GMR_{FM}$ term, especially for the series with 2.0 nm thick Co layers, can be correlated with the reduced surface roughness towards more positive Cu deposition potentials. With reference to Fig. 9, it should be established on the other hand that the assumed structural improvement leading to higher GMR does not show up in the superlattice satellite intensities which are rather reduced for larger Cu deposition potentials.

## Summary

To overcome the deficiencies of previous studies on ED Co/Cu multilayers prepared from a pure sulfate bath, in the present work a systematic study of GMR and structure of such multilayers has been carried out under well-controlled conditions. We described results on samples obtained for a variety of both the magnetic and non-magnetic layer thicknesses as well as the total multilayer thickness in the range from 50 nm to 300 nm. We explored also the influence of Cu deposition potential on the GMR for our ED Co/Cu multilayers.

The XRD measurements revealed superlattice satellite reflections for many of the multilayers having sufficiently high thickness (at least 2.0 nm) of both constituent layers. The bilayer repeats derived from the positions of the visible superlattice reflections were typically 10 - 20% higher than the nominal values.

A clear GMR effect was observed for all multilayers investigated and the GMR was found to be dominated by the multilayer-like $GMR_{FM}$ contribution even for multilayers without visible superlattice satellites. There was always also a modest SPM contribution to the GMR and this term was the largest for multilayers with very thin (0.5 nm) magnetic layers which are already probably not continuous and fully percolating and, thus, contain some amount of small, magnetically decoupled SPM regions as well.

No oscillatory behavior of the $GMR_{FM}$ term with spacer thickness was observed at any magnetic layer thickness. The saturation of the coercivity as measured by the peak position of the MR(H) curves indicated a complete decoupling of magnetic layers for large spacer thicknesses. The measured GMR increased with total multilayer thickness which could be ascribed to an increasing SPM contribution to the GMR due to an increasing surface roughness. This latter behavior could also be used to explain the observed increase of the coercivity.

For multilayers with Cu layers deposited at more and more positive potentials, the $GMR_{FM}$ term increased and the $GMR_{SPM}$ term decreased. At the same time, a corresponding reduction of surface roughness measured with AFM directly indicated an improvement of the multilayer structural quality which was, however, not accompanied by an increase of the superlattice reflections.

The present results, together with previous findings on both electrodeposited and physically deposited multilayers, underline that whereas the structural quality as characterized by the surface roughness generally correlates fairly well with the magnitude of the GMR, the microstructural features determining the amplitude of superlattice reflections apparently do not have directly an influence on the GMR magnitude.

**Acknowledgements** We are indebted to the Hungarian Scholarship Board for providing a 7-month fellowship to N.R. for his research stay in Budapest. This work was supported by the Hungarian Scientific Research Fund (OTKA) through Grant K 104696.

Table 1

| $t_{Co}$ (nm) | 0.5 | 1.1 | 2.0 | 3.0 |
|---|---|---|---|---|
| $t_{Cu}$ (nm) | 4.0 | 4.0 | 4.0 | 4.0 |
| total thickness (nm) | 300 | 300 | 300 | 300 |
| $E_{Cu}$ (mV vs. SCE) | -600 | -600 | -600 | -600 |

Table 2

| $t_{Cu}$ (nm) | 2.0 | 3.0 | 4.0 | 5.0 | 6.0 |
|---|---|---|---|---|---|
| $t_{Co}$ (nm) | 2.0 | 2.0 | 2.0 | 2.0 | 2.0 |
| total thickness (nm) | 300 | 300 | 300 | 300 | 300 |
| $E_{Cu}$ (mV vs. SCE) | -600 | -600 | -600 | -600 | -600 |

Table 3

| $t_{Co}$ (nm) | 1.1 or 2.0 | 1.1 or 2.0 | 1.1 or 2.0 | 1.1 or 2.0 | 1.1 or 2.0 |
|---|---|---|---|---|---|
| $t_{Cu}$ (nm) | 4.0 | 4.0 | 4.0 | 4.0 | 4.0 |
| total thickness (nm) | 50 | 75 | 100 | 200 | 300 |
| $E_{Cu}$ (mV vs. SCE) | -600 | -600 | -600 | -600 | -600 |

Table 4

| $t_{Co}$ (nm) | 1.1 or 2.0 | 1.1 or 2.0 | 1.1 or 2.0 | 1.1 or 2.0 | 1.1 or 2.0 | 1.1 or 2.0 | 1.1 or 2.0 | 1.1 or 2.0 |
|---|---|---|---|---|---|---|---|---|
| $t_{Cu}$ (nm) | 4.0 | 4.0 | 4.0 | 4.0 | 4.0 | 4.0 | 4.0 | 4.0 |
| total thickness (nm) | 300 | 300 | 300 | 300 | 300 | 300 | 300 | 300 |
| $E_{Cu}$ (mV vs. SCE) | -250 | -300 | -350 | -400 | -450 | -500 | -550 | -600 |

Fig. 1 XRD patterns for $[Co(t_{Co})/Cu(t_{Cu})]_{300nm}$ multilayers with Cu layers deposited at $E_{Cu}^{EC}$ = -600 mV (a) with $t_{Cu}$ = 4 nm for various Co layer thicknesses as indicated and (b) with $t_{Co}$ = 2 nm for various Cu layer thicknesses as indicated.

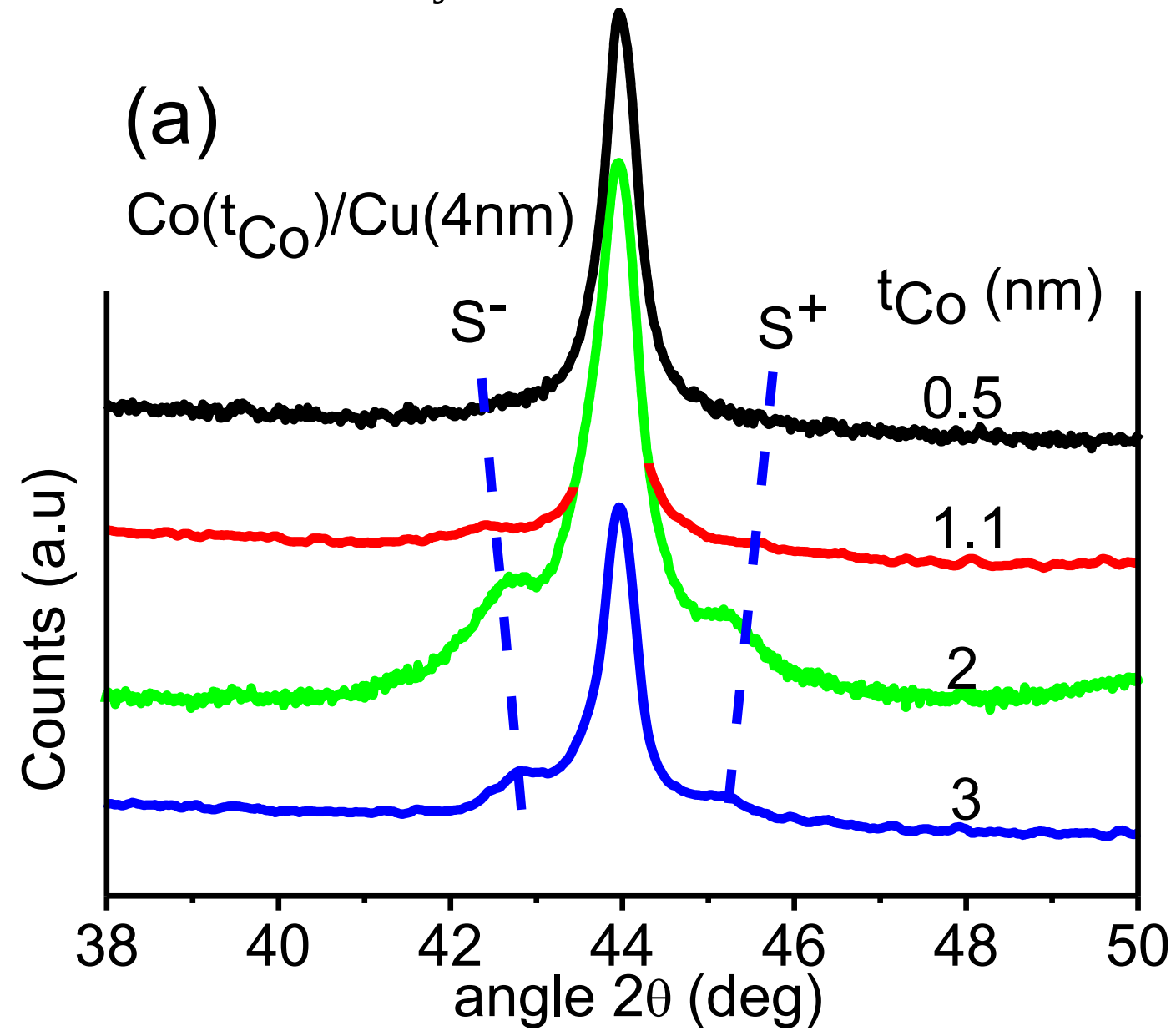

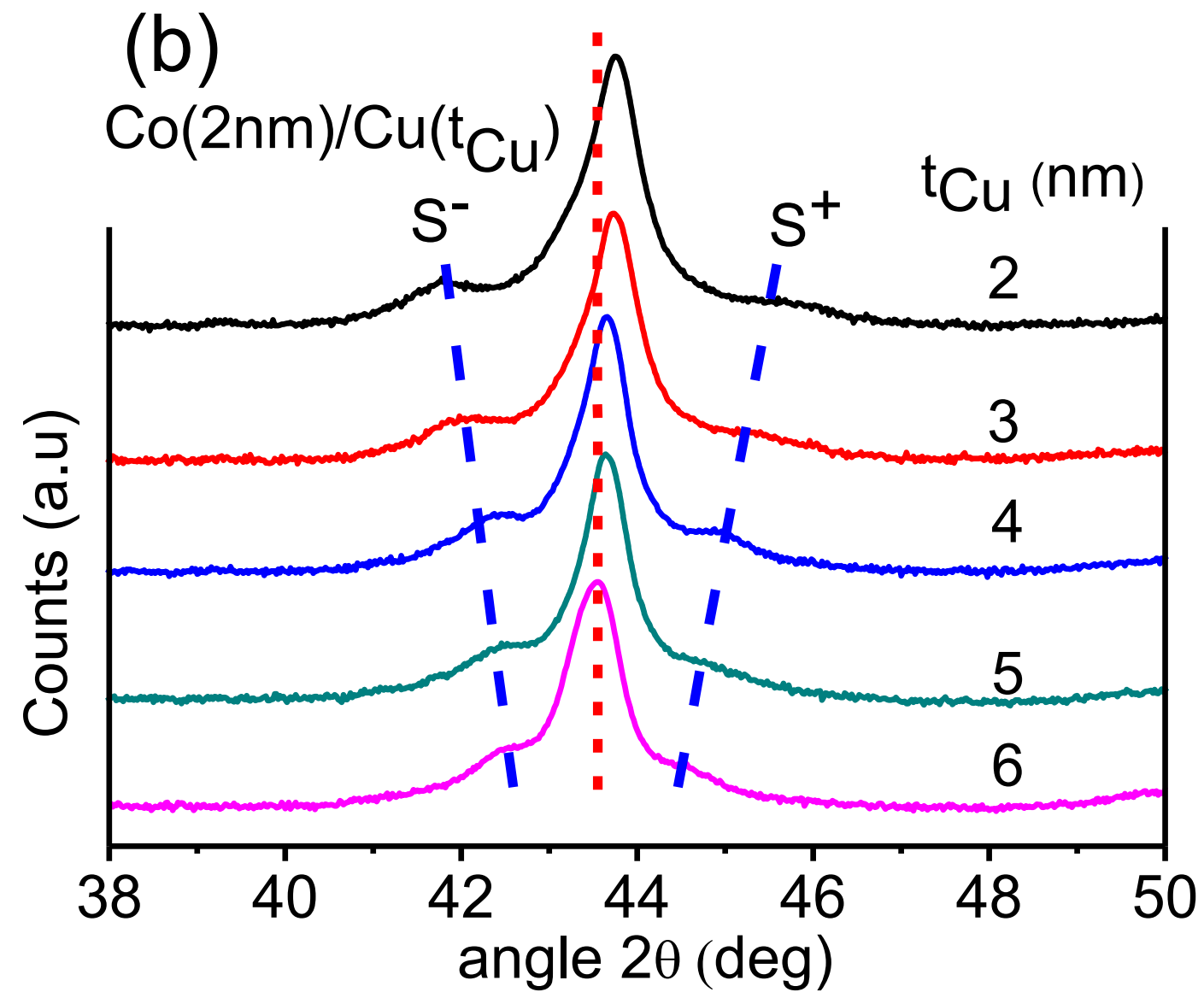

Fig. 2 XRD patterns for (a) [Co(1.1nm)/Cu(4nm)] and (b) [Co(2.0nm)/Cu(4nm)] multilayers with Cu layers deposited at $E_{Cu}^{EC}$ = -600 mV for various total multilayer thicknesses as indicated. The vertical dashed blue lines are intended to indicate that the satellite peak positions if visible at all are expected to remain unchanged with total multilayer thickness since these positions depend on the bilayer length only.

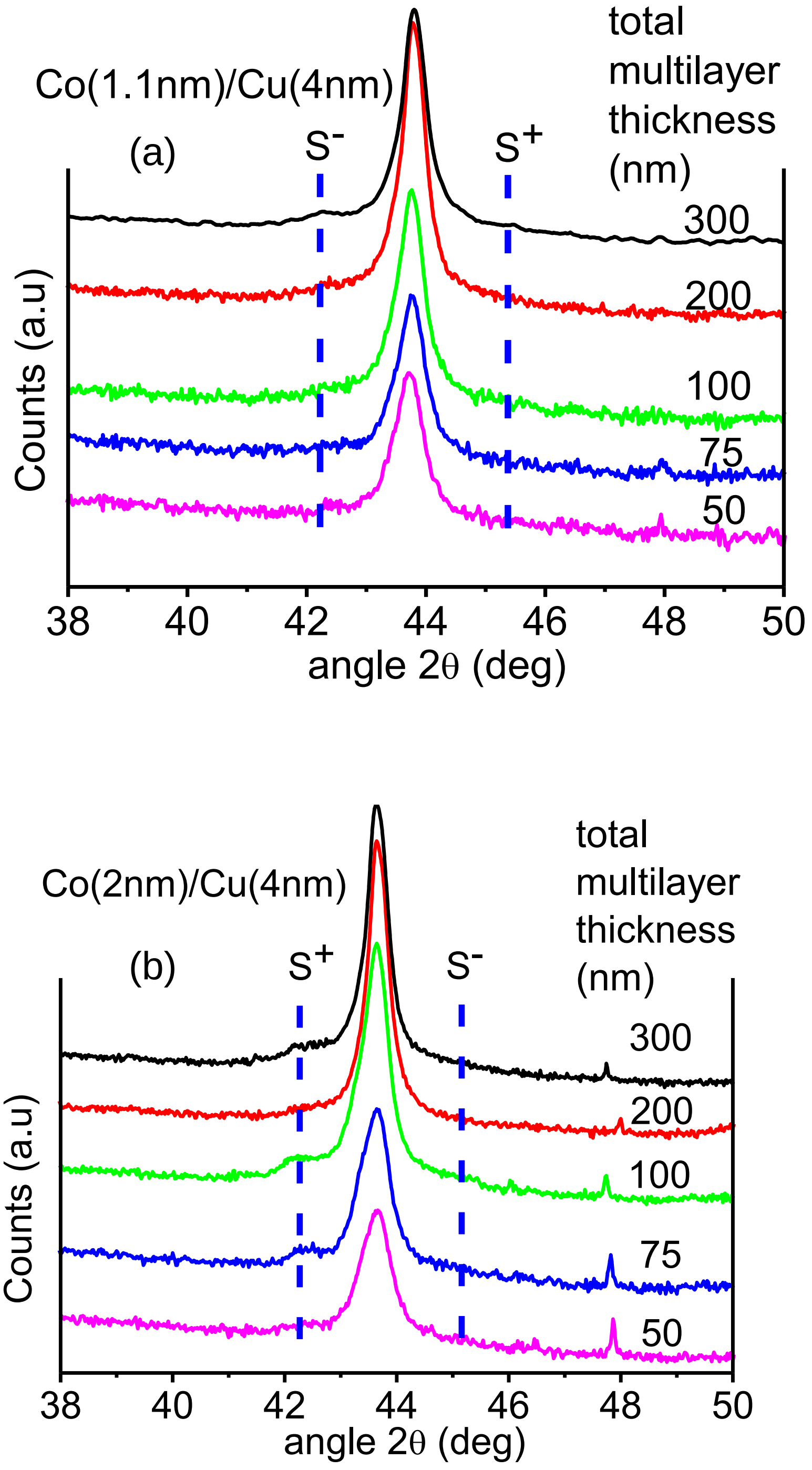

Fig. 3 Bilayer length $\Lambda_{XRD}$ of [Co($t_{Co}$)/Cu($t_{Cu}$)] multilayers with Cu layers deposited at $E_{Cu}^{EC}$ = -600 mV as deduced from the XRD satellite peak positions: (a) $\Lambda_{XRD}/\Lambda_{nom}$ vs. $\Lambda_{nom}$ from the XRD patterns shown in Fig. 1; (b) $\Lambda_{XRD}/\Lambda_{nom}$ vs. total multilayer thickness from the XRD patterns shown in Fig 2. The value of $\Lambda_{nom}$ was obtained by using the nominal layer thicknesses $t_{Co}$ and $t_{Cu}$.

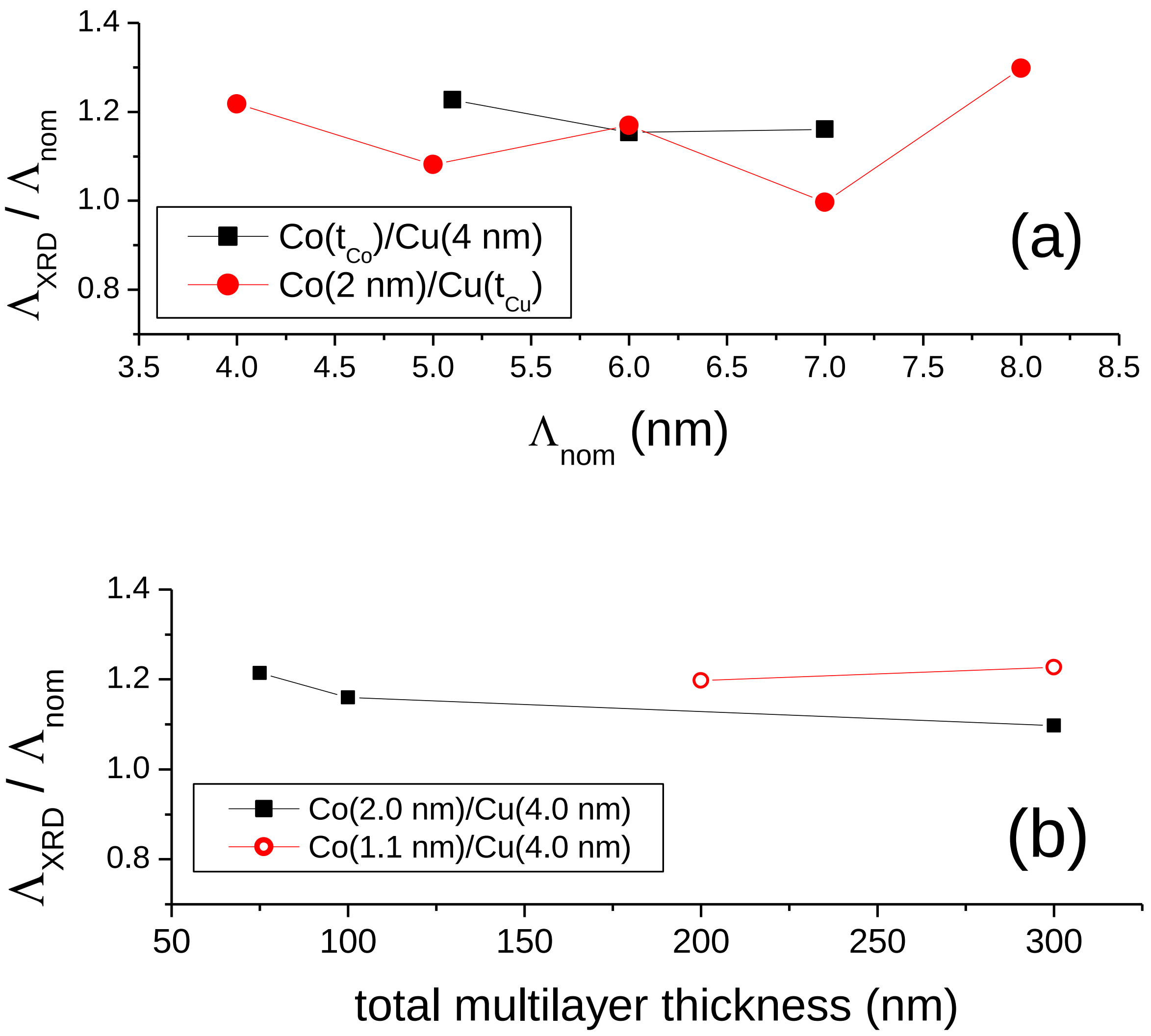

Fig. 4 The two kinds of typical *MR*(*H*) curves measured for Co($t_{Co}$)/Cu($t_{Cu}$) multilayers. Decomposition of the measured longitudinal MR component ($LMR_{Meas}$) into ferromagnetic (FM) and superparamagnetic (SPM) contributions for (a) a Co(0.5nm)/Cu(5nm) and (b) a Co(3nm)/Cu(5nm) multilayer. The decomposition reveals a much larger SPM contribution in the multilayer with very thin magnetic layer.

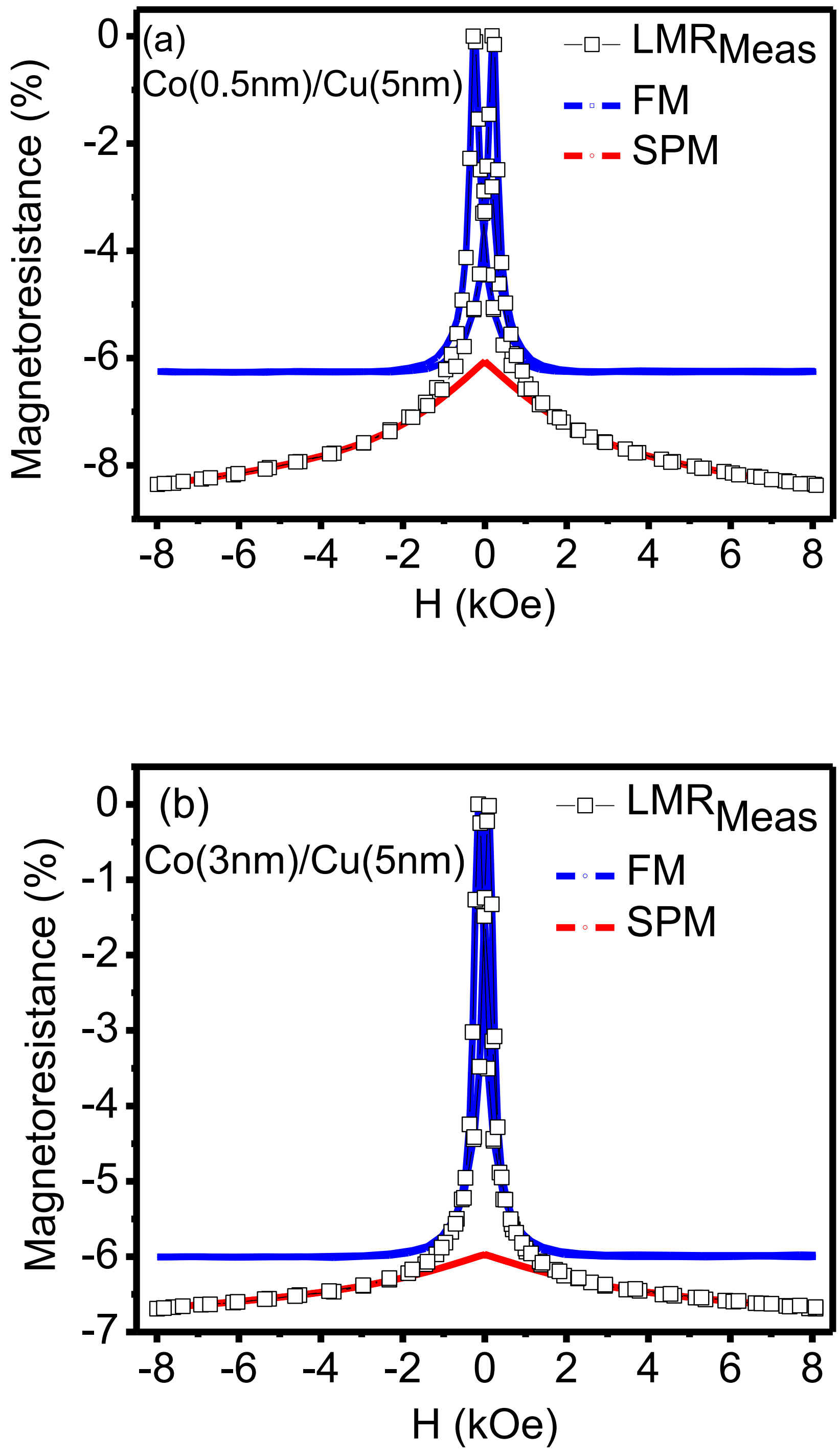

Fig. 5 Magnetoresistance results for $[Co(t_{Co})/Cu(t_{Cu})]_{300nm}$ multilayers with Cu layers deposited at $E_{Cu}^{EC}$ = -600 mV as a function of the Cu layer thickness for various Co layer thicknesses as indicated in the legend. (a) Total saturation $GMR_{is}$ as deduced from the Langevin fitting procedure. (b) Saturation values of the $GMR_{FM}$ and $GMR_{SPM}$ contributions determined by Langevin-fitting performed separately for both the LMR and TMR components; for clarity, their isotropically averaged values are only given. (c) Peak positions ($H_p$) of the measured MR(H) curves. For each sample, an average of the $H_p$ values obtained for the LMR and TMR components is only given.

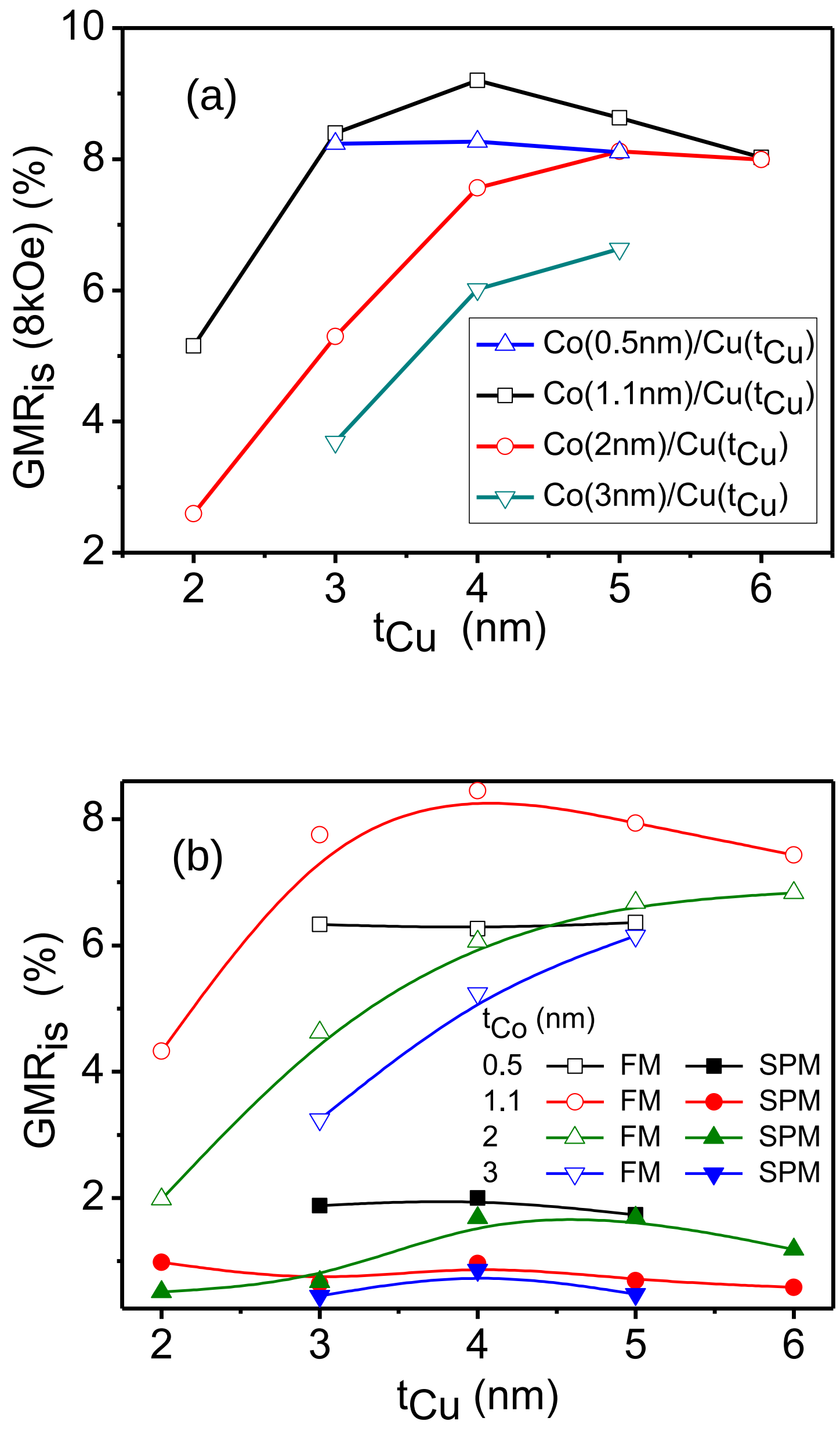

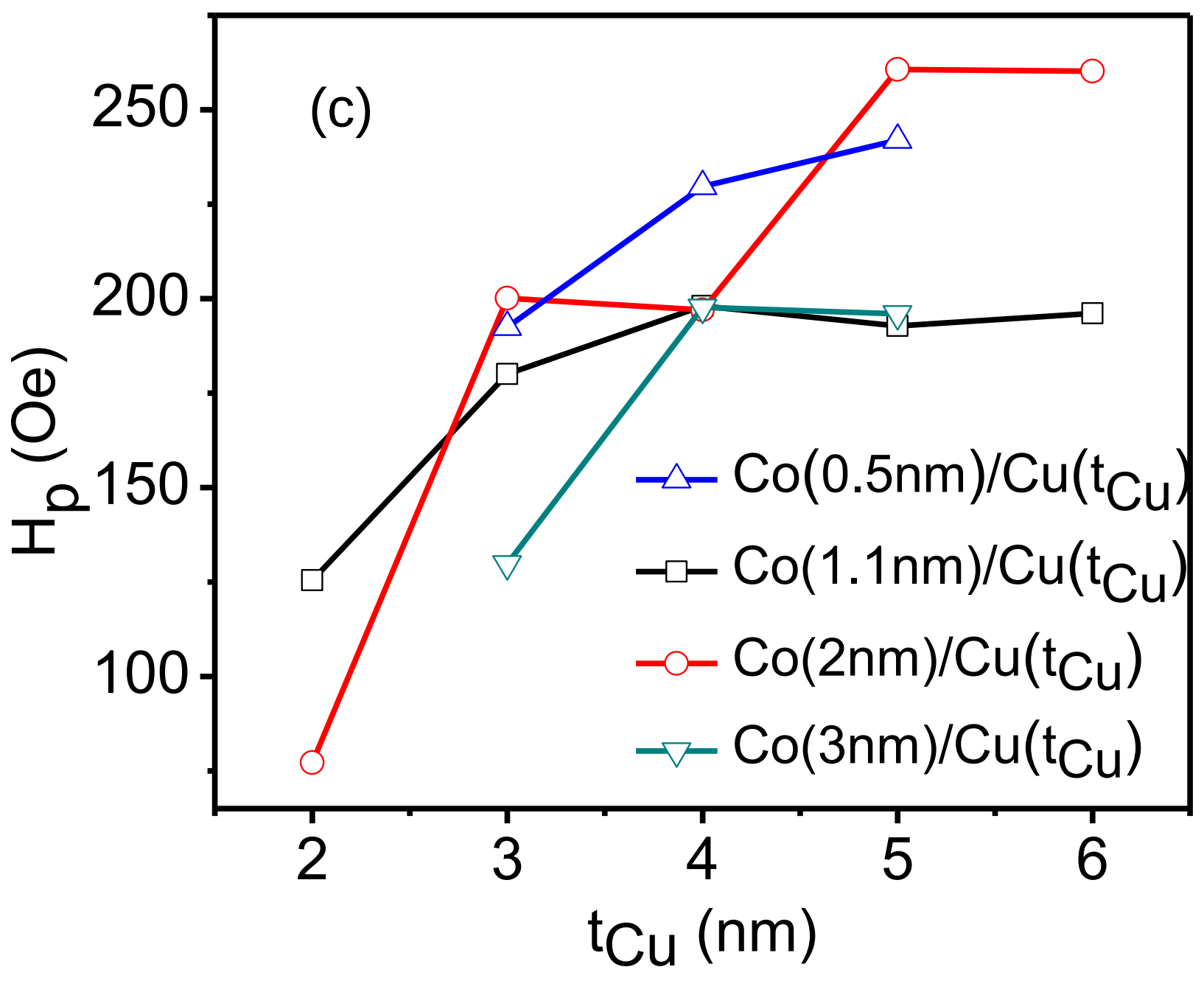

Fig. 6 (a) Total saturation $GMR_{is}$ data as deduced from the Langevin fitting procedure for $[Co(t_{Co})/Cu(4nm)]_{300nm}$ multilayers with Cu layers deposited at $E_{Cu}^{EC}$ = -600 mV as a function of the total multilayer thickness for two Co layer thicknesses as indicated in the legend. (b) The saturation values of the $GMR_{FM}$ and $GMR_{SPM}$ contributions obtained from the Langevin-fitting the MR(H) curves of these multilayers. (c): $H_p$ vs. total multilayer thickness for the two series shown in (a) and (b).

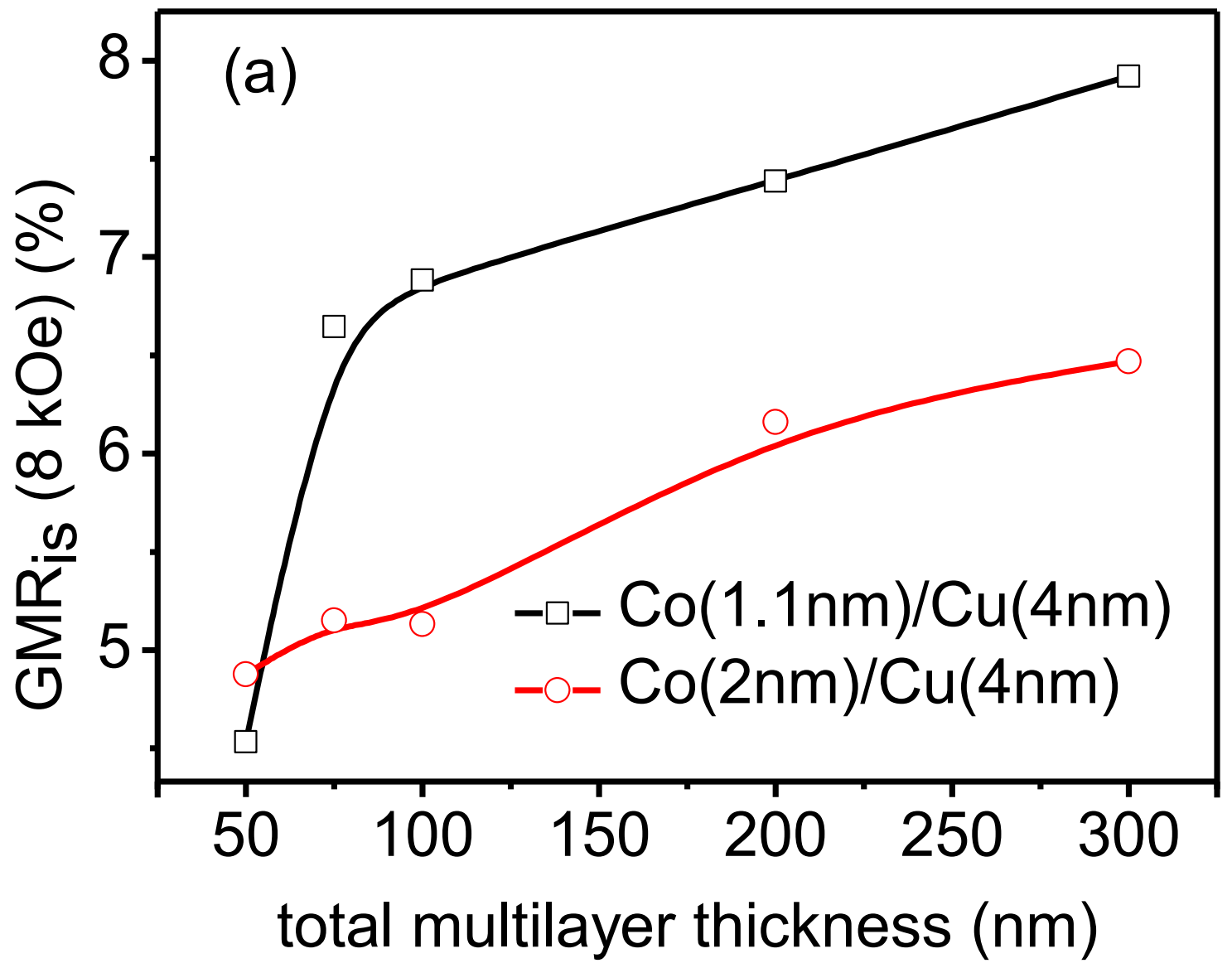

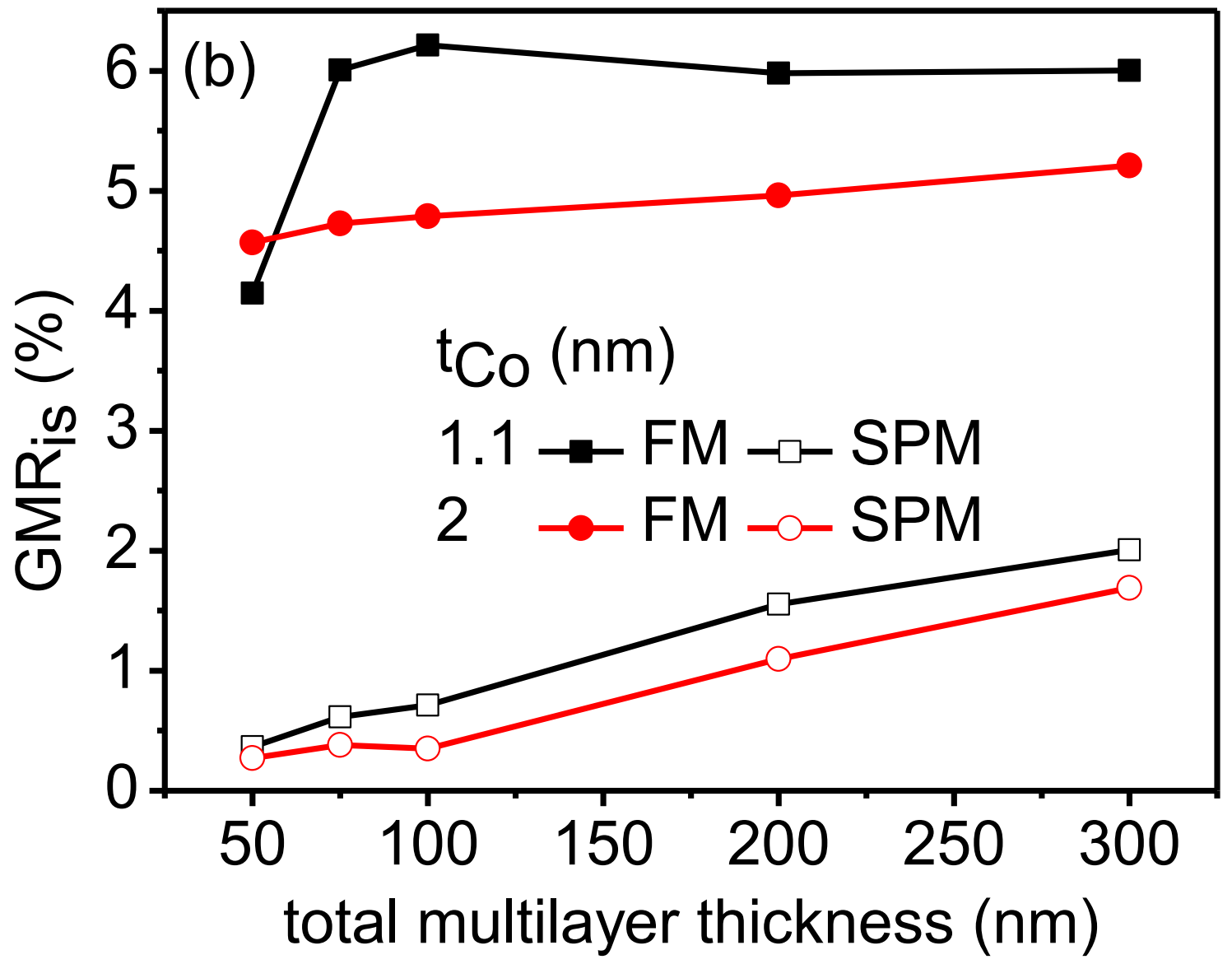

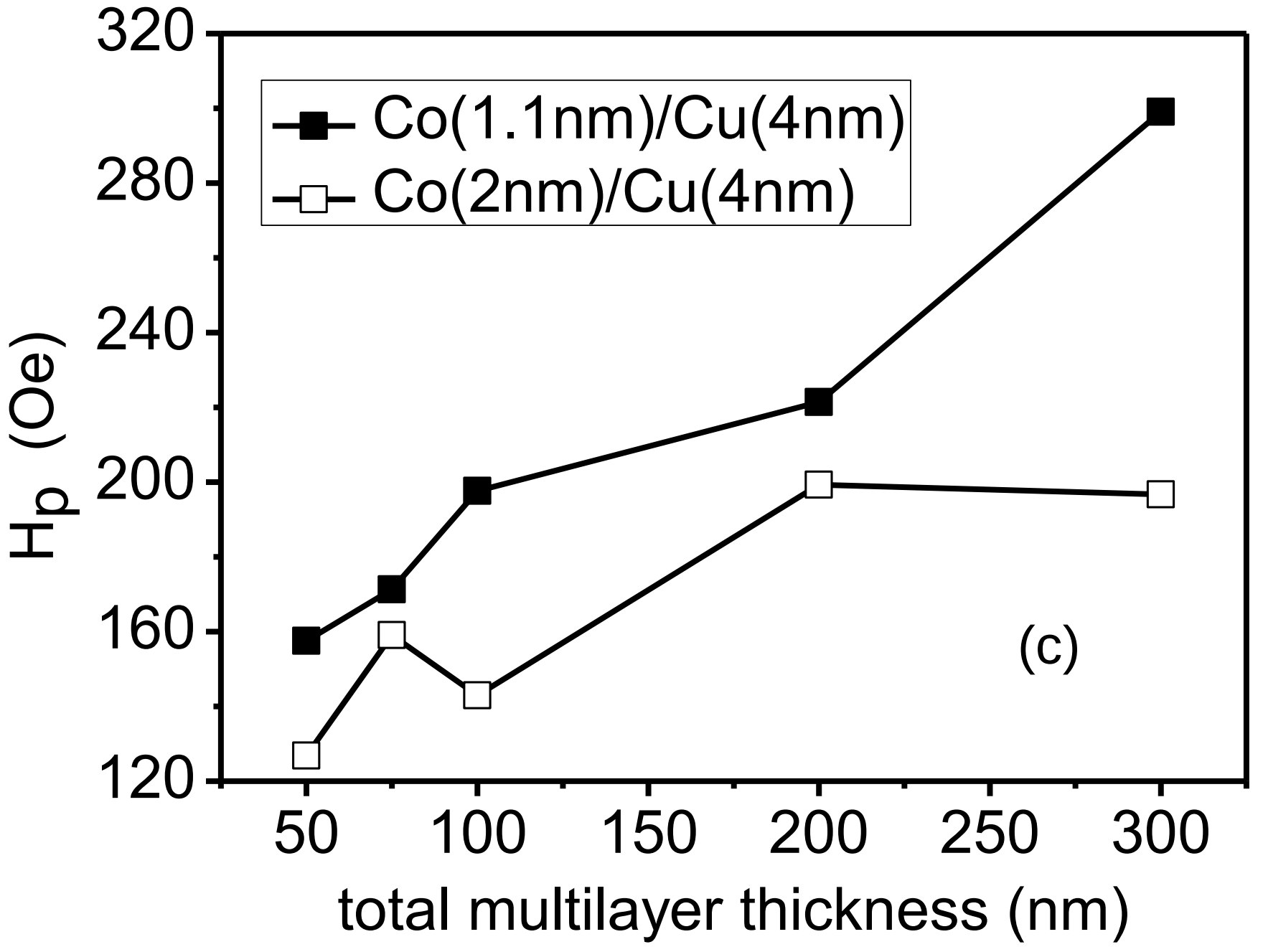

Fig. 7 Overall Cu content from composition analysis of $[Co(1.1nm)/Cu(4nm)]_{300nm}$ and $[Co(2nm)/Cu(4nm)]_{300nm}$ multilayers with Cu layers deposited at the potentials as indicated when measured on their Si/Cr/Cu substrates. Layer thicknesses given are nominal values.

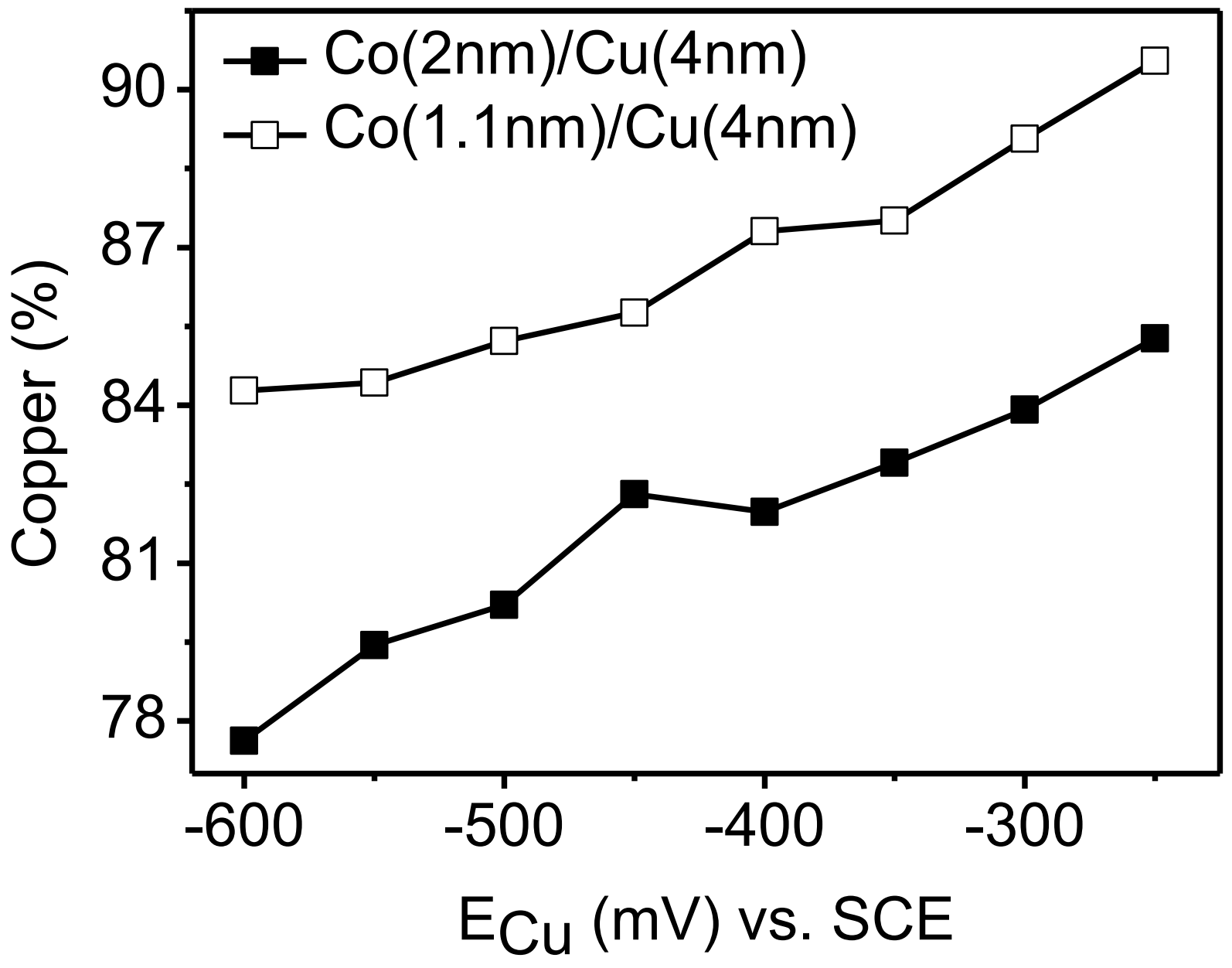

Fig. 8 Evolution of AFM roughness ($R_q$) for $[Co(2nm)/Cu(4nm)]_{300nm}$ multilayers as a function of the Cu layer deposition potential. The dashed line is a guide for the eye only.

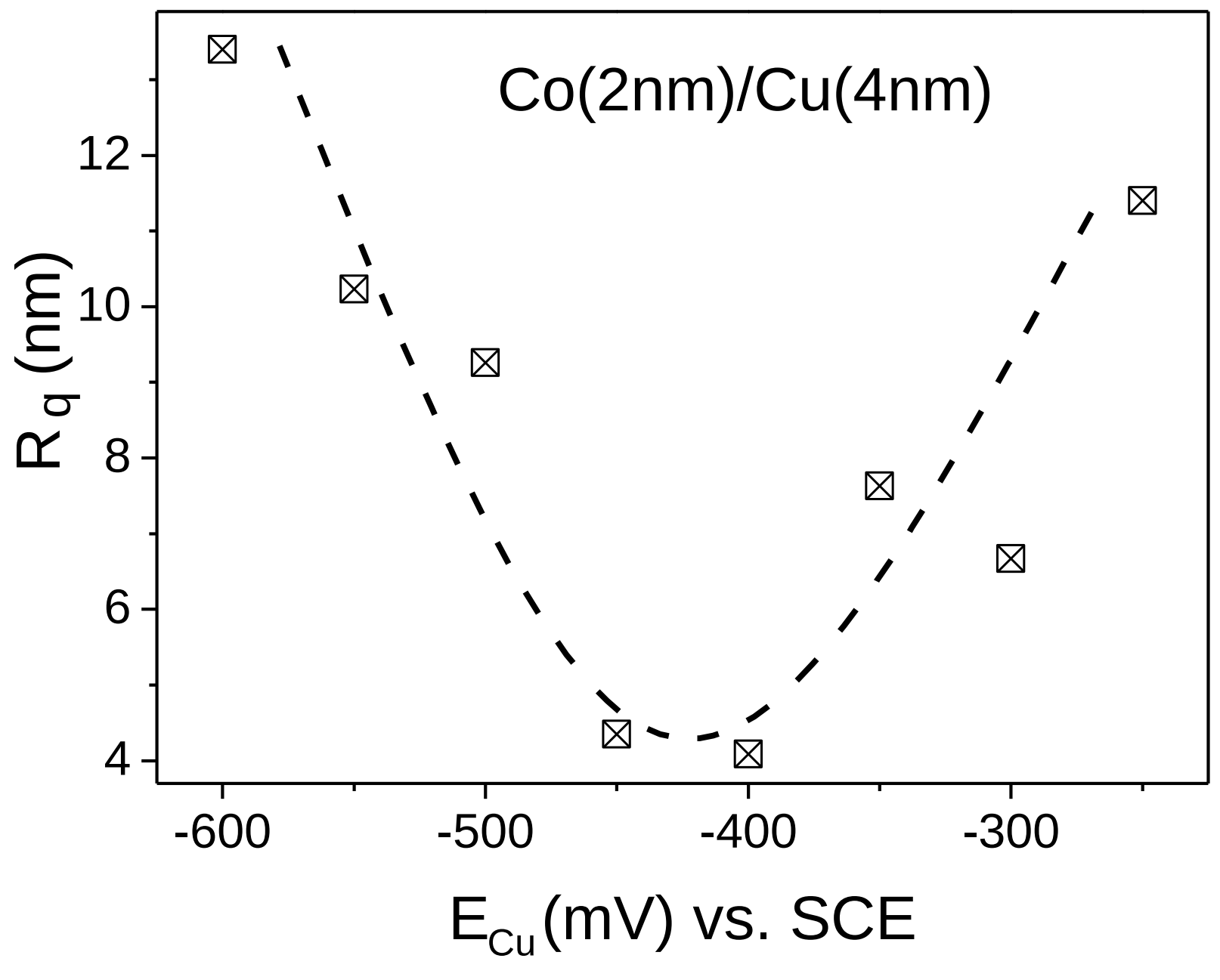

Fig. 9 XRD patterns for (a) $[Co(1.1nm)/Cu(4nm)]_{300nm}$ and (b) $[Co(2nm)/Cu(4nm)]_{300nm}$ multilayers with Cu layers deposited at the potentials as indicated. The small sharp peak around 48 deg in (a) cannot stem from the samples, it is probably a spurious reflection due to the substrate or sample holder. (c) $\Lambda_{XRD}/\Lambda_{nom}$ vs. Cu deposition potential $E_{Cu}$ derived from the XRD patterns shown in (a) and (b). The value of $\Lambda_{nom}$ was obtained by using the nominal layer thicknesses $t_{Co}$ and $t_{Cu}$.

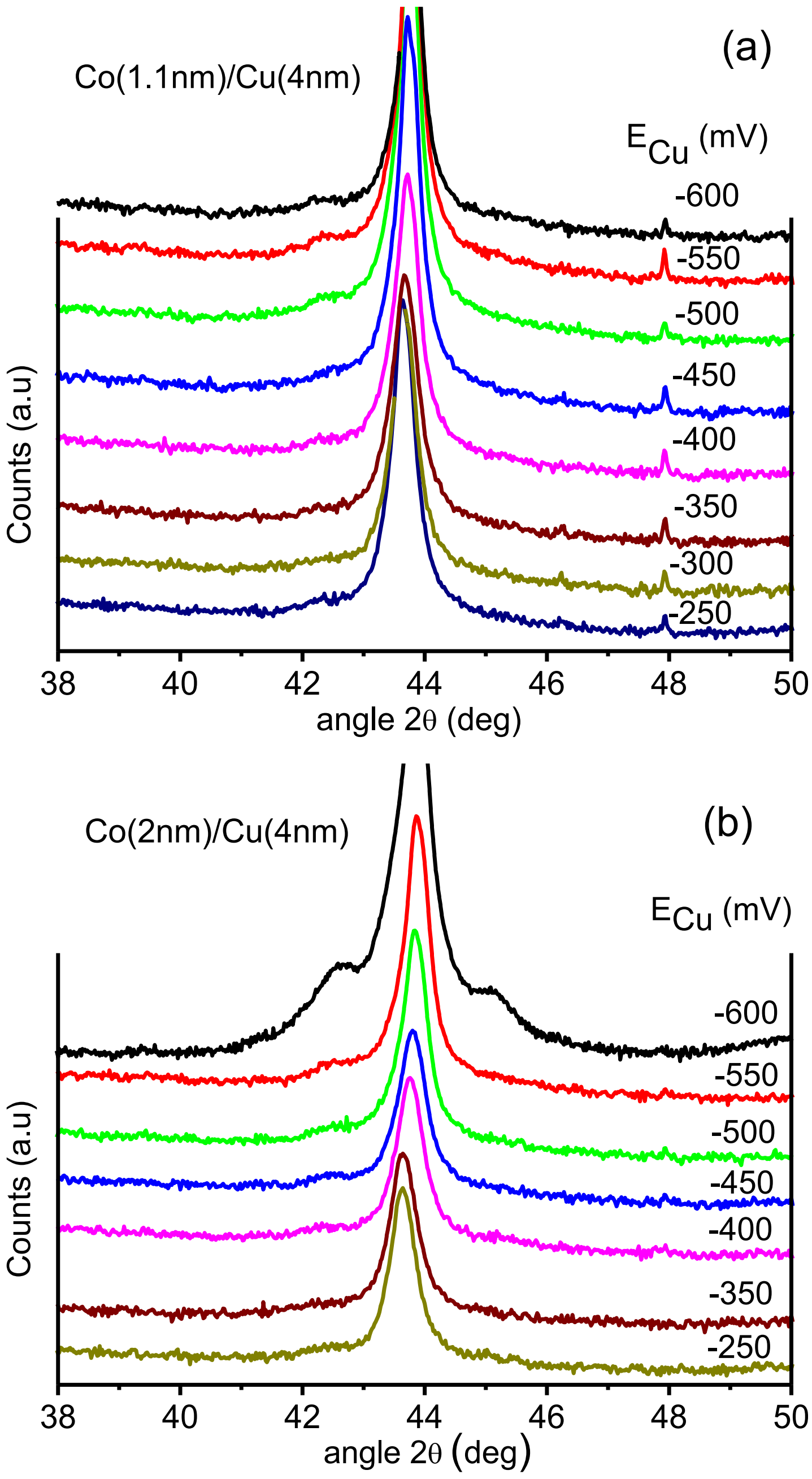

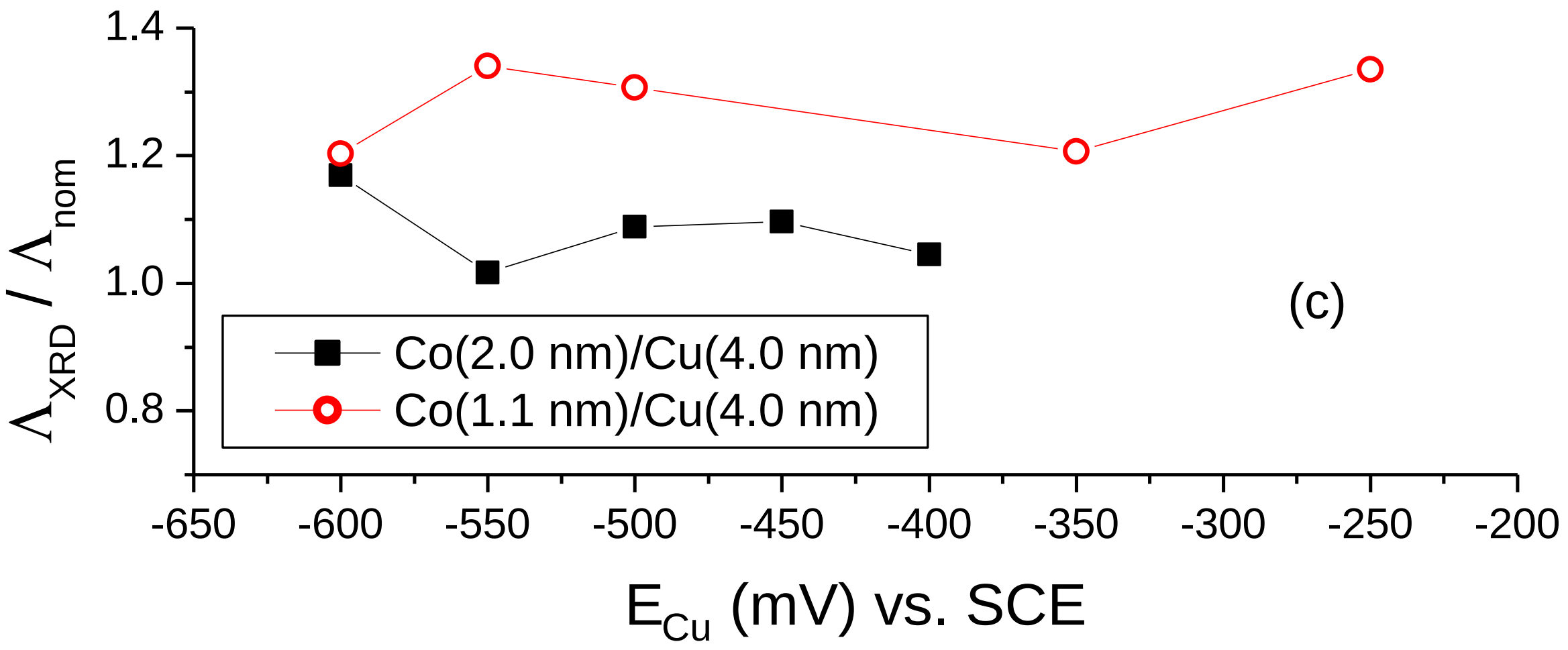

1.4
1.2
1.0
0.8
$\Lambda_{XRD} / \Lambda_{nom}$
(c)
Co(2.0 nm)/Cu(4.0 nm)
Co(1.1 nm)/Cu(4.0 nm)
-650 -600 -550 -500 -450 -400 -350 -300 -250 -200
$E_{Cu}$ (mV) vs. SCE

Fig. 10 (a) Total saturation $GMR_{is}$ data as deduced from the Langevin fitting procedure for $[Co(1.1nm)/Cu(4nm)]_{300nm}$ and $[Co(2nm)/Cu(4nm)]_{300nm}$ multilayers with Cu layers deposited at various $E_{Cu}$ potentials. (b) The decomposed FM and SPM contributions to $GMR_{is}$ for the same multilayers as shown in (a).

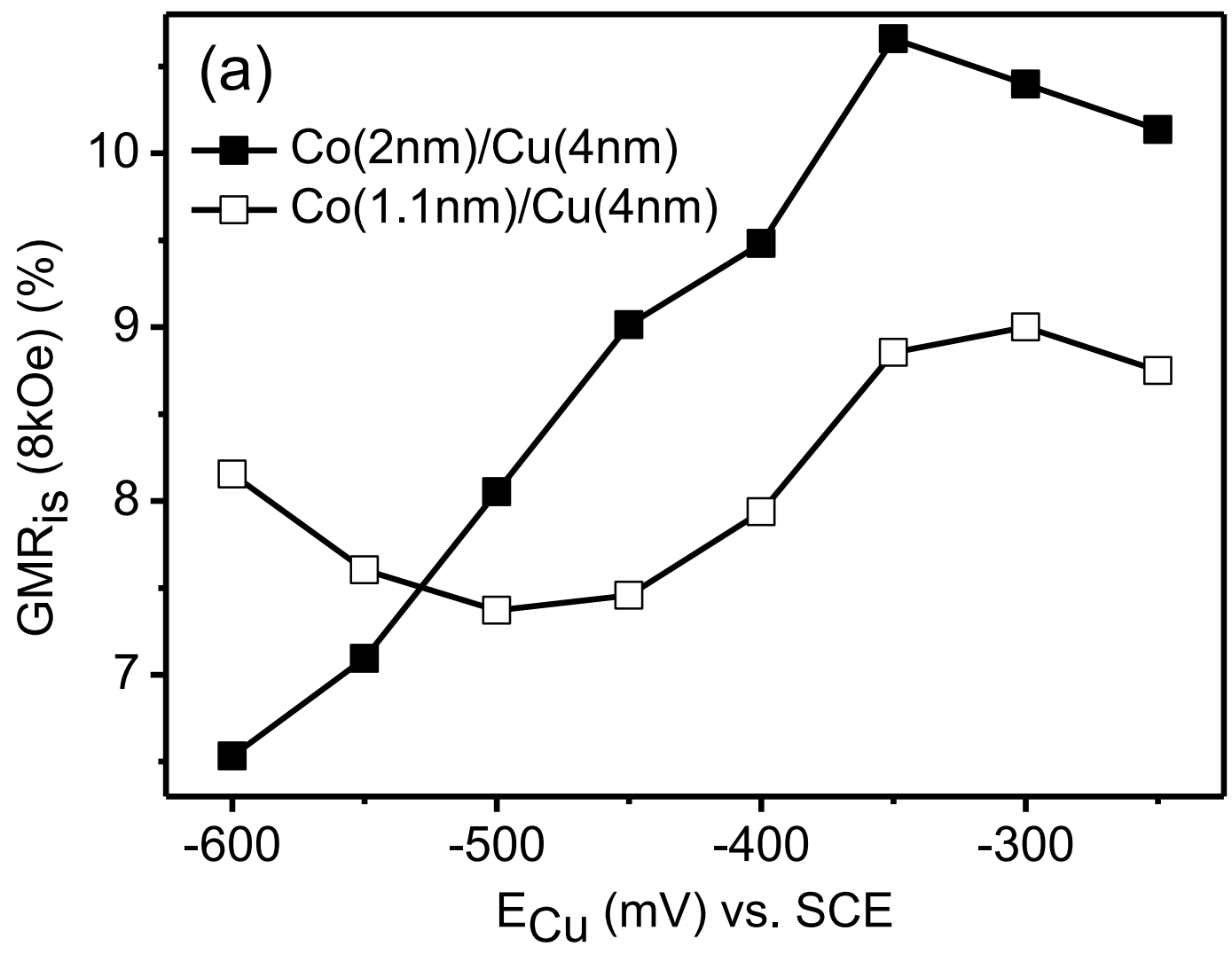

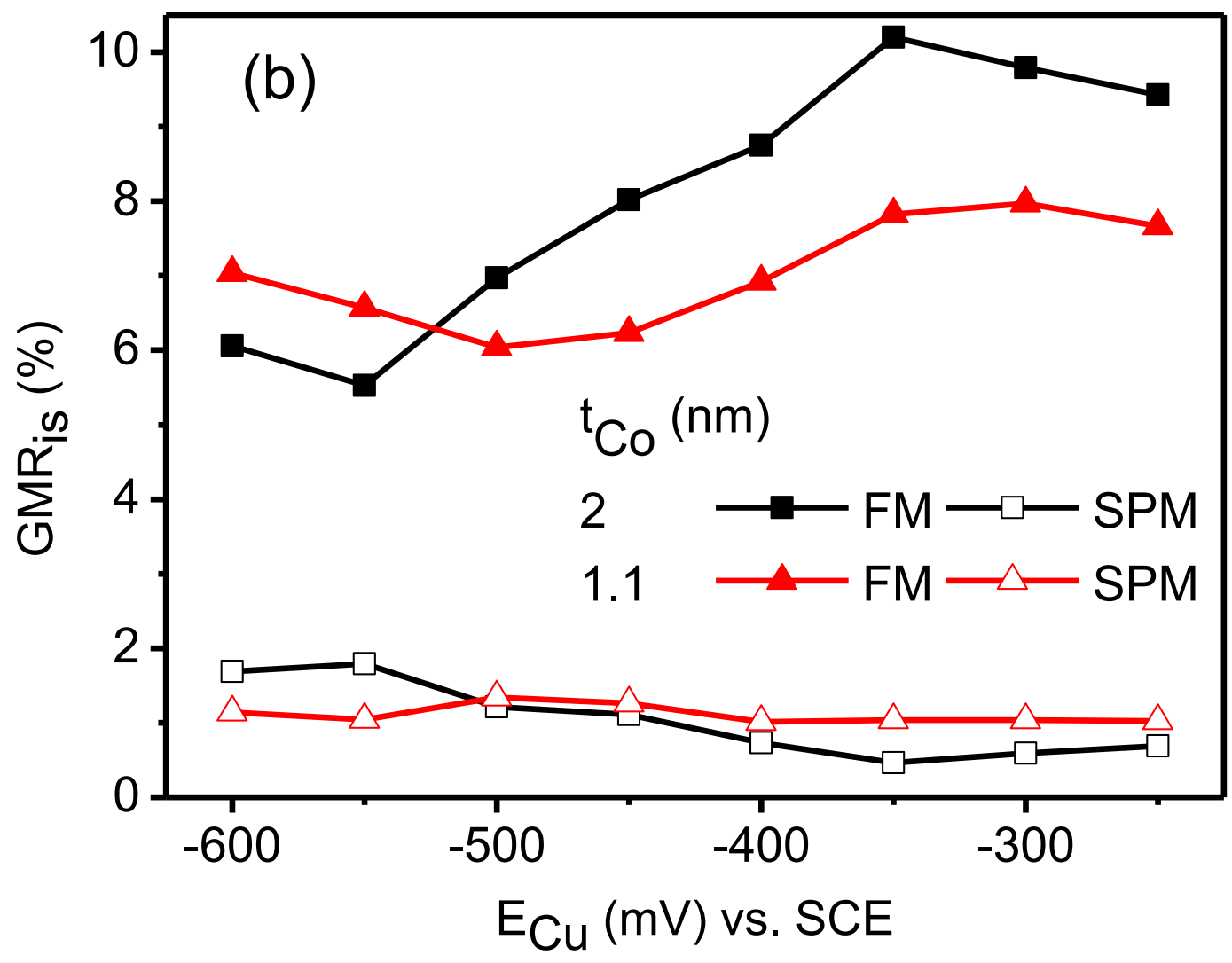